\documentclass[aoas,nameyear,dvips]{arximspdf}
\usepackage{multirow}
\usepackage{graphics}
\doi{10.1214/09-AOAS280}
\volume{4}
\issue{1}
\pubyear{2010}
\firstpage{520}
\lastpage{532}

\begin{document}
\begin{frontmatter}

\title{Nonparametric inference procedure for percentiles of the random
effects distribution in meta-analysis}
\runtitle{Procedure for random effects meta-analysis}

\begin{aug}
\author[a]{\fnms{Rui} \snm{Wang}\corref{}\ead[label=e1]{rwang@hsph.harvard.edu}\thanksref{t1}},
\author[b]{\fnms{Lu} \snm{Tian}\ead[label=e2]{lutian@stanford.edu}\thanksref{t2}},
\author[a]{\fnms{Tianxi} \snm{Cai} \ead[label=e3]{tcai@hsph.harvard.edu}\thanksref{t3}}
\and
\author[a]{\fnms{L. J.} \snm{Wei} \ead[label=e4]{wei@hsph.harvard.edu}\thanksref{t4}}
\address[a]{R. Wang\\
T. Cai\\
L. J. Wei\\
Department of Biostatistics\\
Harvard University School of\\
\quad Public Health\\
Boston, Massachusetts 02115\\
USA\\
E-mail: \printead*{e1}\\
\phantom{E-mail: }\printead*{e3}\\
\phantom{E-mail: }\printead*{e4}}
\address[b]{L. Tian\\
Department of Health Policy\\
\quad and Research\\
Stanford University School\\
\quad of Medicine\\
Stanford, California 94305\\
USA\\
\printead{e2}}
\affiliation{Harvard University, Stanford University, Harvard
University\\ and Harvard University}

\thankstext{t1}{Supported in part by NIH Grants R37 AI24643 and T32 AI007358.}
\thankstext{t2}{Supported in part by NIH Grant R01 HL089778.}
\thankstext{t3}{Supported in part by NIH Grants R37 AI24643 and U54 LM008748.}
\thankstext{t4}{Supported in part by NIH Grants R01 AI052817 and U54 LM008748.}

\runauthor{Wang, Tian, Cai and Wei}
\end{aug}

\received{\smonth{6} \syear{2008}}
\revised{\smonth{8} \syear{2009}}


\begin{abstract}
To investigate whether treating cancer patients with
erythropoiesis-stimulating agents (ESAs) would increase the mortality
risk, Bennett et al. [\textit{Journal of the American Medical Association} \textbf{299} (2008) 914--924]
conducted a meta-analysis with the data
from 52 phase III trials comparing ESAs with placebo or standard of
care. With a standard parametric random effects modeling approach, the
study concluded that ESA administration was significantly associated
with increased \textit{average} mortality risk. In this article we present a
simple \textit{nonparametric} inference procedure for the \textit{distribution} of
the random effects. We re-analyzed the ESA mortality data with the new
method. Our results about the center of the random effects distribution
were markedly different from those reported by Bennett et al. Moreover,
our procedure, which estimates the distribution of the random effects,
as opposed to just a simple population average, suggests that the ESA
may be beneficial to mortality for approximately a quarter of the study
populations. This new meta-analysis technique can be implemented with
study-level summary statistics. In contrast to existing methods for
parametric random effects models, the validity of our proposal does not
require the number of studies involved to be large. From the results of
an extensive numerical study, we find that the new procedure performs
well even with moderate individual study sample sizes.
\end{abstract}

%

\end{frontmatter}
%
\section{Introduction}\label{sec1}
Conventional meta-analysis techniques have been
utilized frequently to make inferences about a single parameter, for
example, the center of the distribution of the random or fixed effects.
Under the random effects model, the procedure for estimating the \textit{mean}
of the random effects proposed by DerSimonian and Laird (DL) (\citeyear{DerSimonian86}) is
routinely used in practice. Their method utilizes a linear combination
of study-specific point estimates with the weights depending on the
within- and among-study variance estimates. This procedure is simple to
implement and does not require patient-level data. Its validity,
however, depends heavily on the individual study sample sizes and the
number of studies [Brockwell and Gordon (\citeyear{Brockwell01}), Bohning et al. (\citeyear{Bohning02}),
Sidik and Jonkman (\citeyear{Sidik07}) and Viechtbauer~(\citeyear{Viechtbauer07})].
In addition, this and other related methods for random effects models
in meta-analysis do not provide inferences about the distribution
function of the random effects. Estimation of this distribution
function or its quantile counterpart provides valuable information for
the complex risk-benefit decision on a new drug or device.

In a meta-analysis using the data from 52 phase III comparative trials
(ESA vs. placebo or standard of care), Bennett et al. (\citeyear{Bennett08}) examined
whether the erythropoiesis-stimulating agents (ESAs) for treating
anemia of cancer patients would increase the patients' risk of
mortality. The point and 95\% interval estimates of two-sample
study-specific hazard ratio were presented in Figure 2 of Bennett et
al. Bennett et al. (\citeyear{Bennett08}) concluded that administration of ESAs was
significantly associated with increased mortality. Using the DL method,
the resulting 95\% confidence interval for the \textit{mean} of the random
hazard ratios (treated vs. untreated with ESA) across the studies was
(1.01, 1.20). The lower bound of the interval is barely above $1$.
Furthermore, it is known that the DL method can produce liberal
confidence interval estimates, that is, the true coverage level tends
to be smaller (sometimes substantially) than the nominal value [Emerson, Hoaglin and
Mosteller~(\citeyear{Emerson93}), Hardy and Thompson (\citeyear{Hardy96}), Brockwell and
Gordon~(\citeyear{Brockwell01,Brockwell07}) and Sidik and Jonkman (\citeyear{Sidik02})]. Therefore, the interval estimates
reported by Bennett et al. may be ``too tight.'' Moreover, from Figure 2
of Bennett et al., it appears that the study-specific hazard ratio
estimates for 22 out of 52 trials are less than $1$, suggesting that
even if the average hazard ratio is more than $1$, the ESA may not be
harmful in all study populations. Last, since the DL method is based on
a weighted average of hazard ratio estimates, the resulting interval
estimates may be sensitive to outliers.

In this article we propose a simple inference procedure for the
percentiles of the random effects distribution based on study-level
data without assuming a parametric form of the distribution. We
re-analyzed the mortality data reported in Bennett et al. (\citeyear{Bennett08}). The
resulting 95\% confidence interval for the \textit{median} of the random
hazard ratios was (0.94, 1.26). The 95\% confidence interval for the
lower quartile of the random hazard ratios was (0.70, 0.99),
indicating that, in approximately a~quarter of the study populations,
ESA treatment may reduce mortality. In contrast to all existing
methods, which can only handle inference for the center of the random
effects distribution, the new proposal does not require the number of
studies to be large. The new proposal is theoretically valid when the
sample sizes of individual studies are large. Through an extensive
numerical study, we find that the new method performs well even with
moderate individual study sample sizes. On the other hand, the DL
method tends to give liberal confidence interval estimators, that is,
their coverage levels can be markedly smaller than the nominal
value.\looseness=1

\section{Interval estimates for percentiles of the random effects
distribution}\label{sec2}
Consider a typical two-level hierarchical model. Let $\Pi
' =(\Theta, \Lambda')$ be a row vector of random parameters, where
$\Theta$ is a univariate parameter of interest and $\Lambda$ is
a~finite- or infinite-dimensional vector of nuisance parameters. Let
$G(\cdot)$ be the continuous, completely unspecified distribution
function of $\Theta.$ Given an \textit{unobservable} realization $\Pi,$ a data
set $X$ is generated. Let $\{\Pi_k, X_k \}, k=1, \ldots, K,$ be $K$
independent copies of $\{\Pi, X\}.$ The problem is how to make
inferences, for instance, about the median $\mu$ of $G(\cdot)$ with $\{
X_k, k=1, \ldots, K\}.$
As an example, consider the case with $K$ $2 \times2$ tables and let
$\Theta_k$ be the log-risk-ratio or risk difference for the $k$th
table. Here, the nuisance parameter $\Lambda_k$ consists of the
underlying event rate for the ``control'' group and the sample size for
the $k$th study $n_k$.

If we can observe $\{\Theta_k, k=1, \ldots, K \},$ a simple
nonparametric estimator for $\mu$ is the sample median. Exact
confidence intervals for $\mu$ can be obtained by inverting a~sign test
for the null hypothesis that the median is
$\mu_0$. Under $H_0\dvtx \mu=\mu_0$, consider
%
%
\begin{equation}\label{eq1}
T(\mu_0)=\sum_{k=1}^K B_k,
\end{equation}
where $B_k=I(\Theta_k < \mu_0)-I(\Theta_k > \mu_0)$ and $I(\cdot)$ is
the indicator function. The null distribution of $T(\mu_0)$ can be
generated by
%
%
\begin{equation}\label{eq2}
T^*= \sum_{i=1}^K \Delta_k   \qquad\mbox{where } \Delta_k= \cases{
\hspace*{8.4pt}1,&\quad  with probability 0.5,\cr
-1,&\quad otherwise.}
\end{equation}

Suppose that, given $\Pi_k,$ $\hat{\Theta}_k$ is a consistent estimator
for $\Theta_k$ based on the data~$X_k$. To test $H_0$, one may replace
$\Theta_k$ in (\ref{eq1}) with $\hat{\Theta}_k.$ This results in the test statistic
%
%
\begin{equation}\label{eq3}
\tilde{T}(\mu_0) = \sum_{k=1}^K \hat{B}_k = \sum_{k=1}^K \{ I(\hat
{\Theta}_k < \mu_0)- I(\hat{\Theta}_k > \mu_0) \}.
\end{equation}
When the sample size $n_k$ for each individual study is large, we can
make inferences about the median by comparing the observed value of (\ref{eq3})
to the distribution of (\ref{eq2}).

Now, the test based on (\ref{eq3}) does not take into account the precision of
the estimator $\hat{\Theta}_k.$ It gives equal weight to each
individual study.
For the $k$th study, suppose that the variance $\hat{\sigma}_k^2$ of
$\hat{\Theta}_k$ is large relative to the distance between $\Theta_k$
and $\mu_0.$ Then the likelihood of the unobservable $\Theta_k < \mu_0
$ can be quite close to $1/2$ (like tossing a fair coin). Therefore, the
noise generated from such an unstable $\hat{B}_k$ may well outweigh its
added value to the power of the test based on $\tilde{T}(\mu_0).$ On
the other hand, if $\hat{\sigma}^2_k$ is small and $\hat{\Theta}_k < \mu
_0$, the likelihood of $\Theta_k < \mu_0 $ would be closer to 1.

This motivates us to modify test statistic (\ref{eq3}) by putting weight $w_k$
on $\hat{B}_k$. Here, $w_k$ is a measure of likelihood of the event
$\Theta_k < \mu_0,$ for example, the observed coverage level of the
interval $(-\infty, \mu_0)$ for the realized $\Theta_k$. When the
individual study size $n_k$ is large, and the distribution of $\hat
{\Theta}_k$ conditional on $\Pi_k$ is approximately normal with mean
$\Theta_k$ and variance $\hat{\sigma}_k^2,$ where $n_k \hat{\sigma
}_k^2$ converges to a constant, this coverage level is approximately
$\Phi((\mu_0-\hat{\Theta}_k)/\hat{\sigma}_k)$, where $\Phi$ is the
distribution function of the standard normal.
Let the resulting test statistic be
%
%
\begin{equation}\label{eq4}
\hat{T}(\mu_0) = \sum_{k=1}^K \bigl|\Phi\bigl((\mu_0 -\hat{\Theta}_k)/\hat{\sigma
}_k\bigr)-1/2\bigr|\hat{B}_k.
\end{equation}

In the \hyperref[app]{Appendix} we show that, in probability, for any given $\mu,$
%
%
\begin{equation}\label{eq5}
\bigl|\Phi\bigl((\mu-\hat{\Theta}_k)/\hat{\sigma}_k\bigr) -1/2\bigr|\hat{B}_k - B_k/2
\rightarrow0\qquad   \mbox{as }  n_k \rightarrow\infty.
\end{equation}
It follows that, for fixed $K,$ for large $n_k, k=1, \ldots, K,$ the
distribution of $\hat{T}(\mu_0)$ approximates that of $T(\mu_0).$ This
approximation, however, is rather discrete; and for moderate sample
sizes, the resulting confidence intervals for $\mu$ do not have
adequate coverage levels in our numerical study (Section \ref{sec4}).
An alternative way to generate an approximation to the null
distribution of $\hat{T}(\mu_0)$ is to use
%
%
\begin{equation}\label{eq6}
\hat{T}^{*}(\mu_0) = \sum_{k=1}^K \bigl|\Phi\bigl((\mu_0 -\hat{\Theta}_k)/\hat
{\sigma}_k\bigr) -1/2\bigr| \Delta_k.
\end{equation}
Here, the $ \Delta_k $'s are the only random quantities and are
analogous to the random multipliers used in the wild bootstrap [Wu
(\citeyear{Wu86})]. The weight from the $k$th study is multiplied by $\Delta_k$,
which is $1$ or $-1$ with probability 0.5 and is generated by the
analyst independently of the observed data. In the \hyperref[app]{Appendix}, we also
justify the asymptotic validity of the test based on (\ref{eq4}) and (\ref{eq6}).
Confidence intervals for~$\mu$ can be obtained by inverting this test.
In contrast to other methods, the new proposal does not require the
number of studies ($K$) to be large. In Section \ref{sec4} we show empirically
that the new interval estimation procedure performs well even when the
sample sizes $(n_k)$ are not large.

The above proposal can be generalized easily to make inferences about
certain percentiles of the distribution $G(\cdot)$. Specifically, let
us hypothesize that the 100$p$th percentile is $\mu_0.$ As for the
median, define $B_k = I(\Theta_k < \mu_0) - I(\Theta_k > \mu_0),$ and
obtain $\hat{B}_k$ by replacing $\Theta_k$ in $B_k$ with $\hat{\Theta
}_k.$ The test statistic is given by
%
%
\begin{equation}\label{eq7}
\hat{T}_p(\mu_0) = \sum_{k=1}^K \bigl|\Phi\bigl((\mu_0 -\hat{\Theta}_k)/\hat
{\sigma}_k\bigr)-1/2\bigr|\hat{B}_k,
\end{equation}
and the null distribution is generated by
%
%
\begin{equation}\label{eq8}
\hat{T}_p^{*}(\mu_0) = \sum_{k=1}^K \bigl|\Phi\bigl((\mu_0 -\hat{\Theta}_k)/\hat
{\sigma}_k\bigr) -1/2\bigr| \Delta_k,
\end{equation}
where $\Delta_k =1$ with probability $p$ and $= -1$ with probability
$1-p.$ Let the resulting test statistic corresponding to (\ref{eq3}) be denoted
by $\tilde{T}_p(\mu_0)$. Confidence intervals for the 100$p$th
percentile can then be obtained by inverting the conditional test accordingly.

\section{Safety meta-analysis of erythropoiesis-stimulating
agents}\label{sec3}
We re-ana\-lyzed the data reported in Bennett et al. (\citeyear{Bennett08})\vspace*{1pt} using the new
proposal. Here $K=52,$ and for the $k$th study,
$\Theta_k$ was the log-hazard ratio and $\hat{\Theta}_k$ was its
estimate. Since the patient-level data were not available, we
approximated the standard error estimate of $\hat{\Theta}_k$ by
one-fourth of the reported length of the 95\% confidence interval
(converted to the log-scale). The 95\% confidence interval for the
median of the distribution of the random hazard ratio $(\exp(\Theta))$
was $(0.94, 1.21)$ based on the test statistic $\hat{T}(\cdot)$ and
(\ref{eq6}). The corresponding interval based on the indicator functions $\{
I(\hat{\Theta}_k < \mu) \}$ via $\tilde{T}(\cdot)$ was
$( 0.90, 1.26),$ which was wider than the above interval. The 95\%
confidence interval for the $mean$ of the random effects distribution
reported in Bennett et al. (\citeyear{Bennett08}) using the DL method was $(1.01,
1.20)$. In the next section we show that the empirical coverage levels
of the DL method can be substantially lower than their nominal
counterparts even when the number of studies is not that small (say, $K=40$).

The 95\% intervals for the $25$th and $75$th percentiles based on (\ref{eq7})
and (\ref{eq8}) were (0.70, 0.99) and (1.18, 1.48), respectively. The
counterparts based on $\tilde{T}_p(\cdot)$ were (0.49, 0.93) and
(1.25, 1.72).
Again, the intervals based on $\hat{T}_p(\cdot)$ were shorter than
those with $\tilde{T}_p(\cdot).$ Note that the upper bound of the 95\%
interval for the 25th percentile was smaller than 1, which suggested
that, approximately, for a quarter of the study populations, their
average hazard ratios for the ESA versus the control were most likely
less than one. That is, on average, the patients in these study
populations may benefit from taking ESA with respect to mortality.

Further investigation to identify characteristics of these trials would
be informative for identifying future cancer patients who would benefit
from the ESAs through reduction of blood cell transfusions and improved
quality of life. On the other hand, it is crucial to identify future
patients who would have unacceptable toxicity risks.

Bennett et al. (\citeyear{Bennett08}) also separately evaluated cancer-related anemia
with six studies (see the top portion of Figure 2 in Bennett et al.)
and investigated whether ESAs would increase the risk of a venous
thromboembolism event (VTE) from~38 comparative phase III trials. The
results obtained using the new proposal are reported in the
supplemental article [Wang et al. (\citeyear{Wang09})].

\section{Numerical studies to evaluate performance of the new
proposal}\label{sec4}
We conducted extensive numerical studies to examine the performance of
the proposed interval estimation procedure for the \textit{percentiles} of
the random effects model under various practical settings. The existing
random effects methods for meta-analysis have focused on making
inferences about the \textit{mean} of the random effects distribution. To
the best of our knowledge, no other methods address the same issue as
our proposed procedure does. Our numerical studies included the DL
interval estimation method, the method proposed by Sidik and Jonkman
(\citeyear{Sidik02}) (SJ), and the one based on $\tilde{T}(\cdot)$ for comparisons.
We considered cases with binary or continuous responses, various
symmetric or asymmetric random effects distributions, and a wide range
of study sample sizes and number of studies. From the results of our
numerical investigation, we find that the new proposal performs well
with respect to the confidence interval coverage level and length. The
DL (or SJ) method tends to be liberal, that is, the empirical coverage
levels can be markedly lower than their nominal counterparts. The
procedure based on the test statistic $\tilde{T}(\cdot)$ produces
confidence intervals whose average lengths are uniformly wider than
those with our method. For percentiles other than the median, the
method based on $\tilde{T}_p(\cdot)$ may have under-coverage.

\begin{figure}[t]

\includegraphics{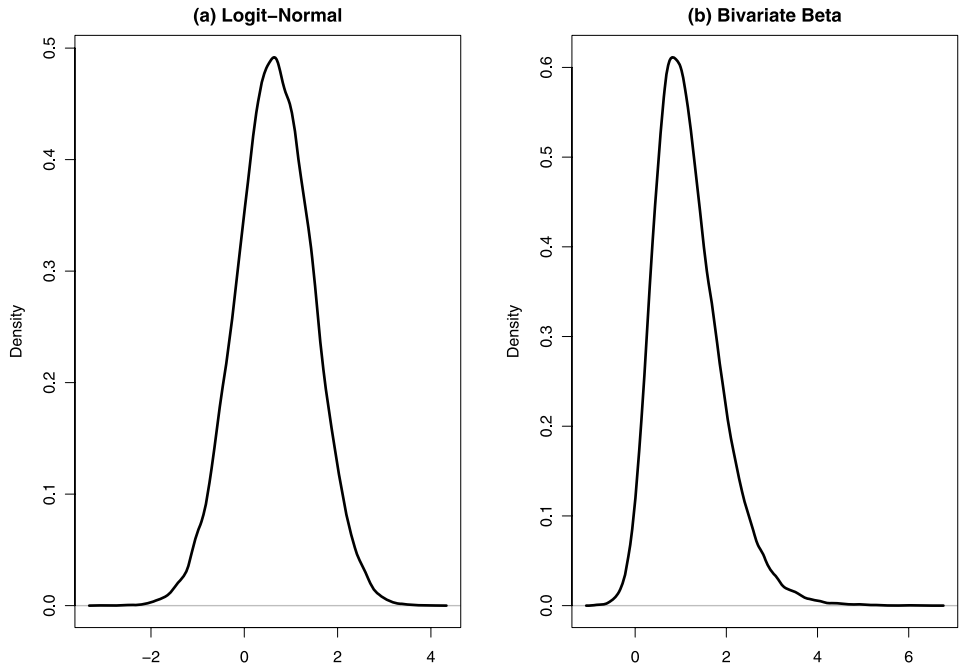}

  \caption{The true density functions for the random log-relative-risk
parameter for the simulation study.}\label{den}\vspace*{-3pt}
\end{figure}

Specifically, in our numerical studies, we first considered
meta-analysis for multiple $2 \times2$ tables under settings similar to
the meta-analysis of VTE rates in Figure~3 of Bennett et al. (\citeyear{Bennett08}).
There are 41 studies listed and the raw data are available for 40
studies. We let $\Theta_k=\log(P_{1k}/P_{0k})$ be the log-relative risk
for the $k$th study, where $P_{1k}$ and $P_{0k}$ are the underlying
event rates for the ESA and control groups, respectively. We then
assumed that the random vectors
$(\operatorname{logit}(P_{0k}), \operatorname{logit}(P_{1k}))'$ were a random sample of
size $K$ from a bivariate normal, whose mean $\eta$ and
variance--covariance matrix $\Sigma$ were estimated by their sample
counterparts via the observed rates in Figure 3 of Bennett et al.
(\citeyear{Bennett08}). We used the conventional 0.5 continuity correction for studies
with zero cells. The resulting sample means and variance--covariance
matrix are $(-3.56, -2.86)'$ and $\left(
{0.90\atop 0.62}\enskip{ 0.62\atop 1.10}\right),$ respectively. The density of $\Theta$ is given in
Figure~\ref{den} [panel (a)], which appears to be quite symmetric. For each
realization $\{(P_{0k}, P_{1k})', k=1, \ldots, K\},$ we generated the
corresponding set of $2 \times2$ tables. We then used DL, SJ, $\hat
{T}(\cdot)$ and $\tilde{T}(\cdot)$ to construct 95\% confidence
intervals for the median of the distribution of $\Theta.$ For each
realized data set, we excluded studies with 0--0 cells (that is, no
events occurred in either group), and used the 0.5 continuity
correction for studies with one zero cell. The average empirical
coverage levels and the median interval lengths were obtained from 2000
realized data sets.\looseness=1

%
\begin{table}[th!]\vspace*{-3pt}
\caption{Empirical coverage levels (ECL) and median lengths (ML) of
95\% interval estimates for median based on DerSimonian--Laird (DL),
Sidik and Jonkman (SJ), $\hat{T}(\cdot)$ and $\tilde{T}(\cdot)$ with a
bivariate logit-normal or a bivariate beta distribution for the two
underlying random event rates}\label{medres}
\begin{tabular*}{\textwidth}{@{\extracolsep{\fill}}lcccccccc@{}}
\hline
\multicolumn{1}{@{}l}{\multirow{2}{50pt}[-5pt]{\textbf{Number of studies,} $\bolds{K}$}} & \multicolumn{2}{c}{\textbf{DL}}& \multicolumn{2}{c}{\textbf{SJ}}&
\multicolumn{2}{c}{$\bolds{\hat{T}(\cdot)}$}& \multicolumn{2}{c@{}}{$\bolds{\tilde{T}(\cdot)}$}
\\[-6pt]
& \multicolumn{2}{c}{\hrulefill}& \multicolumn{2}{c}{\hrulefill}&
\multicolumn{2}{c}{\hrulefill}& \multicolumn{2}{c@{}}{\hrulefill} \\
 & \textbf{ECL}& \textbf{ML} & \textbf{ECL}& \textbf{ML} & \textbf{ECL}& \textbf{ML} & \textbf{ECL}& \textbf{ML} \\
\hline
& \multicolumn{8}{c}{Bivariate logit-normal} \\
$40$ & 86\%& 0.62& 88\%& 0.65 & 94\%& 0.72& 95\%& 0.90 \\
$30$ & 88\%& 0.71& 91\%& 0.75 & 94\%& 0.83& 95\%& 1.03 \\
$20$ & 88\%& 0.85& 91\%& 0.90 & 94\%& 1.00& 95\%& 1.23 \\
$10$ & 88\%& 1.18& 94\%& 1.36 & 95\%& 1.54& 97\%& 2.15 \\
\phantom{0}$6$ &  91\%& 1.57& 97\%& 2.06 & 95\%& 2.29& 97\%& 2.89 \\[6pt]
& \multicolumn{8}{c}{Bivariate beta} \\
$40$& 87\%& 0.40& 89\%& 0.42  &95\%& 0.52& 96\%& 0.65 \\
$30$& 88\%& 0.46& 90\%& 0.48  &95\%& 0.61& 96\%& 0.75\\
$20$& 90\%& 0.55& 92\%& 0.59  &96\%& 0.75& 96\%& 0.91\\
$10$& 91\%& 0.76& 93\%& 0.89  &96\%& 1.10& 98\%& 1.56\\
\phantom{0}$6$& 88\%& 1.00& 94\%& 1.30  &95\%& 1.58& 97\%& 2.10\\
\hline
\end{tabular*}
\end{table}%

Under the same setting, we repeated this process with $K=40$, $30$,
$20$, $10$ and~$6$. For each $K,$ the sample sizes came from the first
$K$ studies listed in Figure 3 of Bennett et al. (\citeyear{Bennett08}). The results are
summarized in Table \ref{medres} (top half). The average coverage
levels for our proposed method, $\hat{T}(\cdot)$, range from 0.94 to
0.95. On the other hand, the average empirical coverage level can be as
low as 0.86 for the DL method, and 0.88 for the SJ method. The median
lengths of the intervals obtained via $\hat{T}(\cdot)$ are uniformly
smaller than those of the procedure using $\tilde{T}(\cdot).$ In
Table~\ref{pctres} (top half), we report the results for the $25$th and
$75$th percentiles. Again our proposal behaves well, but the one with
$\tilde{T}_p(\cdot)$ may not have the correct coverage level.

We also considered rather asymmetric random effects distributions. For
example, we considered a bivariate beta distribution for $\{(P_{0k},
P_{1k})'$, $k=1, \ldots, 40\}$ via three independent gamma random
variables with a common unit scale parameter and shape parameters of
$2$, $8$ and $10$, respectively [Olkin and Liu (\citeyear{Olkin03})]. The resulting
density function of the random parameter $\Theta,$ the log-relative
risk, is given in Figure \ref{den} [panel (b)].
Under the same setting as the previous simulation, the results are
reported in the bottom half portions of Tables \ref{medres} and \ref{pctres}.
Again, the new procedure performs well. The DL (or SJ) method
still has coverage problems. Although the DL method produces confidence
interval estimates for the mean of $G(\cdot)$, not the median, its
empirical coverage for the mean was also lower than the nominal 95\%.
For example, when $K=40$, the coverage of DL for the mean was only 64\%.

%
\begin{table}
\caption{Empirical coverage levels (ECL) and median lengths (ML) of
95\% confidence intervals for the 25th and 75th percentiles based on
$\hat{T}_p(\cdot)$ and $\tilde{T}_p(\cdot)$ with a bivariate
logit-normal or a bivariate beta distribution for the two underlying
random event rates}\label{pctres}
\begin{tabular*}{\textwidth}{@{\extracolsep{\fill}}lcccccccc@{}}
\hline
& \multicolumn{4}{c}{\textbf{25th percentile}} & \multicolumn{4}{c@{}}{\textbf{75 percentile}}
\\[-6pt]
& \multicolumn{4}{c}{\hrulefill} & \multicolumn{4}{c@{}}{\hrulefill} \\
\multicolumn{1}{@{}l}{\multirow{2}{50pt}[-5pt]{\textbf{Number of studies,} $\bolds{K}$}}& \multicolumn{2}{c}{$\bolds{\hat{T}_p(\cdot)}$} & \multicolumn{2}{c}{$\bolds{\tilde{T}_p(\cdot)}$} &
\multicolumn{2}{c}{$\bolds{\hat{T}_p(\cdot)}$}& \multicolumn{2}{c@{}}{$\bolds{\tilde{T}_p(\cdot)}$} \\[-6pt]
& \multicolumn{2}{c}{\hrulefill} & \multicolumn{2}{c}{\hrulefill} &
\multicolumn{2}{c}{\hrulefill}& \multicolumn{2}{c@{}}{\hrulefill} \\
& \textbf{ECL}& \textbf{ML}& \textbf{ECL}& \textbf{ML}& \textbf{ECL}& \textbf{ML}& \textbf{ECL}& \textbf{ML} \\
\hline
&\multicolumn{8}{c}{Bivariate logit-normal} \\
$40$& 95\%& 0.86& 86\%& 1.16 & 95\%& 0.81& 92\%& 0.92 \\
$35$& 96\%& 0.91& 88\%& 1.21 & 96\%& 0.86& 90\%& 1.02 \\
$30$& 96\%& 1.00& 90\%& 1.37 & 96\%& 0.94& 91\%& 1.12 \\
$25$& 96\%& 1.12& 90\%& 1.49 & 97\%& 1.06& 92\%& 1.23 \\
$20$& 96\%& 1.24& 92\%& 1.52 & 97\%& 1.16& 92\%& 1.32 \\[6pt]
&\multicolumn{8}{c}{Bivariate beta} \\
$40$& 96\%& 0.48& 93\%& 0.55 & 96\%& 0.73& 92\%& 0.96 \\
$35$& 96\%& 0.52& 95\%& 0.61& 96\%& 0.78& 93\%& 1.04 \\
$30$& 95\%& 0.56& 94\%& 0.64 & 96\%& 0.85& 93\%& 1.07\\
$25$& 96\%& 0.62& 93\%& 0.65& 96\%& 0.94& 92\%& 1.10 \\
$20$& 96\%& 0.72& 95\%& 0.80& 96\%& 1.37& 95\%& 1.37\\
\hline
\end{tabular*}
\end{table}

Although our method assumes that the random effects distribution is
continuous, we also considered cases with fixed effects models in our
numerical study. For example, we let $(P_{0k}, P_{1k})=(0.1, 0.2), k=1,
\ldots, K.$ The results are summarized in Table \ref{fixres}. For this
case, the DL method has correct coverage level for most scenarios under
which our interval estimation procedure is comparable with the DL
method with respect to efficiency, which is reflected in the interval length.
We also studied the performance of our method for $\Theta_k= P_{1k} - P_{0k},$
the risk difference for the $k$th study. The results were very similar
to those for the relative risk.\looseness=1

%
\begin{table}
\caption{Empirical coverage levels (ECL) and median lengths (ML) of
95\% interval estimates for median based on DerSimonian--Laird (DL),
$\hat{T}(\cdot)$ and $\tilde{T}(\cdot)$ under a fixed effect model (the
underlying event rates are 0.1 and 0.2)}\label{fixres}
\begin{tabular*}{\textwidth}{@{\extracolsep{\fill}}lcccccc@{}}
\hline
 \multicolumn{1}{@{}l}{\multirow{2}{50pt}[-6pt]{\textbf{Number of studies,} $\bolds{K}$}}
 & \multicolumn{2}{c}{\textbf{DL}}& \multicolumn{2}{c}{$\bolds{\hat{T}(\cdot
)}$}& \multicolumn{2}{c@{}}{$\bolds{\tilde{T}(\cdot)}$} \\[-6pt]
 & \multicolumn{2}{c}{\hrulefill}& \multicolumn{2}{c}{\hrulefill}& \multicolumn{2}{c@{}}{\hrulefill} \\
& \textbf{ECL}& \textbf{ML} & \textbf{ECL}& \textbf{ML} & \textbf{ECL}& \textbf{ML} \\
\hline
$40$ & 92\%& 0.24& 95\%& 0.27& 96\%& 0.35 \\
$30$ & 94\%& 0.26& 95\%& 0.30& 96\%& 0.39 \\
$20$ & 95\%& 0.30& 95\%& 0.35& 97\%& 0.45 \\
$10$ & 97\%& 0.47& 96\%& 0.57& 98\%& 0.84 \\
\phantom{0}$6$ & 96\%& 0.75& 95\%& 1.03& 97\%& 1.34\\
\hline
\end{tabular*}
\end{table}

Our numerical studies with continuous responses yielded similar
results. We summarize the study settings and the results in the
supplemental article [Wang et al. (\citeyear{Wang09})]. We expect similar results
for censored time to event observations, where hazard ratios are used
for treatment effect measurements.

\section{Discussion}\label{sec5}
In this article we present a simple nonparametric interval estimation
procedure for percentiles of the random effects distribution. Random
effects meta-analysis is frequently employed in medical research.
However, the \mbox{validity} of the most popular method (DL) and its
variations [Hardy and Thompson (\citeyear{Hardy96}), Biggerstaff and Tweedie (\citeyear{Biggerstaff97}),
Hartung (\citeyear{Hartung99}), Hartung and Knapp (\citeyear{Hartung01a,Hartung01b}) and DerSimonian and
Kacker (\citeyear{DerSimonian07})] is not clear when the number of studies is not large or
the parametric assumption for the random effects is violated. An
excellent review on meta-analysis with the random effects model is
given by Sutton and Higgins (\citeyear{Sutton08}). In contrast to previous methods,
our proposal does not require the number of studies to be large. The
new proposal is valid provided the individual study sample sizes are large.

In addition, if the random effects distribution is symmetric and the
$exact$ distribution of $\hat{\Theta}_k,$ $k=1, \ldots, K,$ conditional
on $\Pi_k$, is symmetric around the unknown fixed realized $\Theta_k,$
it is easy to show that the resulting interval estimators based on $\hat
{T}(\cdot)$ for the median (or mean) are valid without requiring the
sizes of the individual studies or the number of studies to be large.
For instance, under the usual two-sample location shift model with
continuous response variable, let $\Theta$ be the location shift
parameter of interest. Then, the two-sample rank estimator $\hat{\Theta
}$ is symmetric around $\Theta$ under rather mild conditions [Lehmann
(\citeyear{Lehmann75}), page 86]. If the unspecified random effects distribution is
symmetric around $\mu,$ one can use our procedure to obtain exact
confidence intervals for $\mu.$ To examine the performance of the
method in this setting, we conducted a simulation study, described in
detail in the supplemental article [Wang et al. (\citeyear{Wang09})].

The proposed procedure can be implemented with study level summary
statistics. When patient level data are available, various novel
procedures have been studied for mixed effects regression models for
continuous, discrete or censored event time observations [Laird and
Ware (\citeyear{Laird82}), Hougaard (\citeyear{Hougaard95}), Hogan and Laird (\citeyear{Hogan97}),
Henderson, Diggle and Dobson~(\citeyear{Henderson00}), Lam, Lee and Leung (\citeyear{Lam02}), Nelder, Lee and Pawitan (\citeyear{Nelder06}), Cai,
Cheng and Wei (\citeyear{Cai02}), Zeng and Lin (\citeyear{Zeng07}) and Zeng, Lin and Lin
(\citeyear{Zeng08})]. To the best of our knowledge, all of the existing asymptotic
procedures for mixed effects models assume that the number of studies
is large.

In the current practice of meta-analysis, inferences are made only for
the ``center'' of the random effects distribution. A conclusion on the
risk or benefit from an intervention based solely on an estimated
center of the random effects distribution provides limited information
and is usually not sufficient. If the number of studies involved is not
small, we highly recommend estimating this distribution or its
percentiles as proposed in this article.

Under the fixed effects model, this distribution has a single unknown
mass point. The standard estimation procedure for such a fixed
parameter value utilizes a weighted average of study-specific point
estimates. For analyzing multiple $2 \times2$ tables, the most
commonly used procedures are the Mantel--Haenszel [Mantel and Haenszel
(\citeyear{Mantel59})] and Peto methods [Yusuf et al. (\citeyear{Yusuf85})]. These methods are valid
when the number of studies and each individual study sample size are
large. Moreover, when the event rate is small, these standard methods
may not perform well. For the fixed effects model, Tian et al. (\citeyear{Tian09})
proposed a general exact interval estimation procedure that combines
study-specific exact confidence intervals instead of point estimates.
If the fixed effects model is approximately correct, the existing
interval procedures for the common parameter value $\mu$ may be more
efficient than those developed under the random effects model. The
standard heterogeneity tests generally do not have the power to detect
violations of the fixed effects modeling assumption. Therefore, in
practice, sensitivity analyses with both random and fixed effects
models are highly recommended.

\appendix

\section*{Appendix: Justification for the conditional
test $\hat{T}(\cdot)$ based on the approximation generated by
$\hat{T}^*(\cdot)$}\label{app}

Let $D_k = |\Phi((\mu-\hat{\Theta}_k)/\hat{\sigma}_k)-1/2|\hat{B}_k -
B_k/2$. We show that $D_k$ goes to~0, in probability,
as $n_k \rightarrow\infty.$ Here, the probability is generated by the
random element $(X_k,\Pi_k).$ For any fixed positive constant $c,$
first we show that \mbox{$\operatorname{pr}(|D_k|\ge c |  \Pi_k) \rightarrow0$} for any
given $\Pi_k$ with $\Theta_k \neq\mu.$ To this end, consider two cases.
First, if $\Theta_k< \mu,$ then conditional on $\Pi_k,$
\[
|D_k|=\bigl|\Phi\bigl((\mu-\hat{\Theta}_k)/\hat{\sigma}_k\bigr)-1\bigr|=1-\Phi\bigl((\mu-\Theta
_k)/\hat{\sigma}_k+(\Theta_k-\hat{\Theta}_k)/\hat{\sigma}_k\bigr).
\]
As $n_k \to\infty$,
$ (\mu-\Theta_k)/\hat{\sigma}_k \rightarrow\infty$ in
probability, and
$ (\Theta_k-\hat{\Theta}_k)/\hat{\sigma}_k\rightarrow N(0,1)$
in distribution.
Therefore, for any $c>0$, we can find $N$
such that, when \mbox{$n_k>N$},
$\operatorname{pr} ( (\mu-\Theta_k)/\hat{\sigma}_k+ (\Theta_k-\hat{\Theta
}_k)/\hat{\sigma}_k\le\Phi^{-1}(1- c )) < c,$ which is equivalent to
$\operatorname{pr}(\Phi((\mu-\hat{\Theta}_k)/\hat{\sigma}_k ) <1- c)=\operatorname{pr}(|D_k|\ge c )< c.$ Therefore,
$\operatorname{pr}(|D_k|\ge c \mid\Pi_k) \rightarrow0$. Similarly, if
$\Theta_k > \mu$, we can show that $\operatorname{pr}(|D_k| \ge c \mid\Pi_k)
\rightarrow0$ as $n_k\rightarrow\infty.$ Therefore,
$\operatorname{pr}(|D_k|\ge c \mid\Pi_k) \rightarrow0$ for any $\Pi_k$ such
that $\Theta_k \ne\mu.$

This, coupled with the
fact that $G(\cdot)$ is continuous,
implies that $\operatorname{pr}(|D_k| \ge c) = \mbox{E}_{\Pi_k}\{\operatorname{pr}(|D_k| \ge c \mid\Pi_k) \} \rightarrow0 $ for any
$c$ by the dominated convergence theorem. Therefore,
$D_k \to0$ in probability as $n_k\rightarrow\infty.$
It follows that $|\hat{T}(\mu) -\sum_{k=1}^K B_k/2| \to0,$
in probability, as $\min\{n_1,\ldots,n_K\}\rightarrow\infty.$

Similarly, since
\[
\bigl|\bigl|\Phi\bigl((\mu-\hat{\Theta}_k)/\hat{\sigma}_k\bigr)-1/2\bigr|\Delta_k -
|I(\Theta_k < \mu)-1/2|\Delta_k \bigr|\le|D_k|,
\]
one can show that $\hat{T}^*(\mu) - \sum_{k=1}^K |I(\Theta_k < \mu
)-1/2|\Delta_k\to0,$
in probability, as $\min\{n_1,\ldots,n_K\}\rightarrow\infty,$
where
\[
\Delta_k= \cases{
\hspace*{8.4pt}1,&  \quad $\mbox{with probability } p,$ \cr
-1,& \quad $\mbox{with probability } 1-p,$}
\]
for the 100$p$th percentile and is independent of the data.
Therefore,
for any $t$ and positive $c$,
\begin{eqnarray*}
&&\operatorname{pr}_{\{(X_k, \Pi_k)_{k=1,\ldots,K}\}}\Biggl(\Biggl|\operatorname{pr} \bigl(\hat
{T}^*(\mu) \le t|(X_k,\Pi_k)_{k=1,\ldots,K}\bigr)-
\operatorname{pr} \Biggl( \sum_{k=1}^K \Delta_k/2\le t\Biggr)\Biggr|\ge c \Biggr)\\
&&\qquad \le c,
\end{eqnarray*}
when $\min\{ n_1, \ldots, n_K \}$ is large.
This, coupled with the fact that $\sum_{k=1}^K B_k/2 \sim\sum
_{k=1}^K\Delta_k/2$ under the
null hypothesis that the 100$p$th percentile of $\Theta_k$ is $\mu$, implies
that one can approximate the null distribution of $\hat{T}(\mu)$ by
the distribution of $\hat{T}^*(\mu)$ conditional on the observed data.

\section*{Acknowledgments}
We thank Professor Michael Newton, Dr. David Hoaglin and a referee for
their comments, which improved the paper.
{\sloppy
\begin{supplement}
\stitle{Additional examples, simulation results and computer codes}
\slink[doi]{10.1214/09-AOAS280SUPP}
\slink[url]{http://lib.stat.cmu.edu/aoas/280/supplement.pdf}
\sdatatype{.pdf}
\sdescription{We present the results for the mortality data set
restricted to the six trials for anemia of cancer and the results for
the venous thromboembolism rates data set in Bennett et al. (\citeyear{Bennett08})
using the proposed approach, report the simulation results for
continuous responses and for the setting where the sample sizes for
individual studies are small, and provide R codes for implementation of
the proposed procedure.}
\end{supplement}}

\printaddresses

\end{document}